\documentclass[final,twocolumn]{elsarticle}
\usepackage{graphicx}
\usepackage{graphics}
\usepackage{amssymb}

\begin{document}

\begin{frontmatter}

\title{Gravitational collapse in pure Lovelock gravity in higher dimensions}

\author[nkd1,nkd2]{Naresh Dadhich}
\ead{nkd@iucaa.ernet.in}

\author[nkd1]{Sushant G. Ghosh}
\ead{sgghosh2@jmi.ac.in}

\author[nkd1]{Sanjay Jhingan\corref{cor1}}
\ead{sanjay.jhingan@gmail.com}

\address[nkd1]{Centre for Theoretical Physics, Jamia Millia Islamia,
New Delhi 110025, India}

\address[nkd2]{Inter-University Centre for
Astronomy \& Astrophysics, Post Bag 4, Pune 411 007, India}

\cortext[cor1]{Corresponding author}

 \begin{abstract}
We study collapse of inhomogeneous dust and null dust (Vaidya radiation) in pure Lovelock gravity in higher dimensions. Since pure Lovelock  gravity is kinematic in odd $d=2N+1$ dimension, hence pertinent dimension for the study is even $d=2N+2$, where $N$ is degree of Lovelock polynomial. It turns out that pure Lovelock collapse favors naked singularity as against black hole for the Einstein case in the same dimension while strength of singularity as measured by divergence of Kretshmann scalar is interestingly the same in the two cases; i.e. the corresponding scalars have the same fall off behavior.
 \end{abstract}

\end{frontmatter}


\section{Introduction}
Lovelock Lagrangian is a homogeneous polynomial of degree $N$ in Riemann curvature and it defines a general action for gravitation. Einstein-Hilbert action is included as linear order $N=1$ while Gauss-Bonnet is quadratic, $N=2$. The characteristic and distinguishing feature of Lovelock action is, notwithstanding its  polynomial character, that the resulting equation of motion for gravitation is always second order. This is however well-known. We wish to point out one other distinguishing universal property of Lovelock gravity (all along by Lovelock gravity we would mean pure Lovelock $N$th order term without previous terms $<N$). As Einstein gravity is kinematic in $3$ dimension and becomes dynamic in $4$ dimension, Lovelock gravity, as defined by $N$th order Riemann curvature, $R^{(N)}_{abcd}$ which is a homogeneous polynomial in Riemann, is also similarly kinematic in $d=2N+1$ and becomes  dynamic in $d=2N+2$ dimension for any $N$ \cite{dgj}. That is $R{(N)}_{ab} = 0$ implies  $R^{(N)}_{abcd} =0$ in all odd $2N+1$ dimensions, in particular $N=2$ Gauss-Bonnet is  kinematic in $5$ and becomes dynamic in $6$ dimension. This is also a universal feature of Lovelock gravity.

The natural question that arises is, what should be the right gravitational equation for higher dimension $d>4$? Should it be Einstein or Einstein-Lovelock or pure Lovelock in $d=2N+1, 2N+2$ \cite{dgj, d1}? If kinematic and dynamic character in odd and even dimension is to be a universal feature of gravity, it cannot be anything else than Lovelock gravity. This property rules out Einstein and Einstein-Lovelock gravity for $d>4$ and singles out Lovelock gravity. Further it also turns out that for Lovelock gravity in odd and even dimensions, thermodynamical parameters bear a universal relation to horizon radius; i.e. entropy always goes as $r_h^2$ in even $d=2N+2$ dimension \cite{dpp1}. It is both necessary and sufficient condition for universality of thermodynamical parameters in terms of horizon radius that it is Lovelock gravity.

It has been proposed \cite{n1} that right gravitational equation in $d>4$ is pure Lovelock equation which includes Einstein equation for $N=1$. In this paper, we shall in particular investigate problem of gravitational collapse which concerns an important open question, what is the ultimate fate of a collapsing star, is it a black hole or naked singularity? Penrose proclaimed in his famous Cosmic Censorship Conjecture (CCC) that end result of collapse is always a black hole. That is singularity is always covered by event horizon and there occur no naked singularities. There do however exist examples of inhomogeneous dust collapse in which there do emanate null geodesics from singularity before apparent horizon is formed thereby indicating nakedness of singularity in general relativity \cite{psj}, Einstein Gauss Bonnet \cite{sgsj}, and Einstein Lovelock gravity \cite{ohashi}. It would be interesting to study this problem for pure Lovelock gravity and contrast it with Einstein gravity in higher dimensions. It would be shown that collapse dynamics would be qualitatively similar for $d = 2N+1,  2N+2$ for Lovelock gravity to that of Einstein gravity in $3, 4$ dimensions. In contrast it would be quite different for Einstein gravity in higher dimensions. This would be yet another discriminator between Einstein and Lovelock gravity.

The paper is organised as follows. In the next Sec. we give a quick overview of Lovelock gravity which is followed by study of gravitational collapse of dust and Vaidya null radiation. We conclude with a discussion.

\section{Lovelock Gravity - a quick overview}

Following Dadhich \cite{bianchi} we define the Lovelock curvature polynomial
\[
R^{(N)}_{abcd}= F^{(N)}_{abcd}-\frac{n-1}{n(d-1)(d-2)} F^{(N)} (g_{ac}g_{bd} - g_{ad}g_{bc}),
\]
\[
F^{(N)}_{abcd} = Q_{ab}{}{}^{mn} R_{cdmn} ,
\]
where
\[
Q^{ab}{}{}_{cd} = \delta^{a b a_1 b_1...a_{N-1} b_{N-1}}_{cdc_1 d_1...c_{N-1}
d_{N-1}} R_{a_1 b_1}{}{}^{c_1 d_1}...R_{a_{N-1} b_{N-1}}{}{}^{c_{N-1} d_{N-1}},
\]
and
\[
Q^{abcd}{}{}{}{}_{;d}=0.
\]
The analogue of $N^{th}$ order Einstein tensor is given by
\[
G^{(N)}_{ab} = N(R^{(N)}_{ab} - \frac{1}{2} R^{(N)} g_{ab}) ,
\]
and
\[
 R^{(N)} = \frac{d-2N}{N(d-2)}F^{(N)} .
\]
Note that $R^{(N)}=R^{(N)}_{ab}g^{ab}=0$ in $2N$ dimension for
arbitrary metric $g_{ab}$. Since $R^{(N)}_{ab}$ is a function of the
metric and its first and second derivatives which are all arbitrary,
it must vanish in $d=2N$. That is, $R^{(N)}_{ab}=0$ identically in
$2N$ dimension. On the other hand for the general Lovelock case, the
lagrangian is non-zero for $d=2N$ but its variation vanishes
identically.

The Lovelock equation of motion is given by
 \begin{equation}\label{equation}
 G^{(N)}_{ab} = -\Lambda g_{ab} + \kappa T_{ab}.
 \end{equation}

\section{Gravitational Collapse}

We study gravitational collapse of dust and null dust in Lovelock gravity in higher dimensions. As expected it would turn out that for Lovelock collapse the situation would be qualitatively the same in odd $2N+1$ and even $2N+2$ as for the Einstein case in $3, 4$ dimensions. In contrast, it would be quite different for Einstein gravity in higher dimensions.

\subsection{Inhomogeneous dust Collapse}

Spherically symmetric homogeneous dust collapse model analysed by Oppenheimer and Snyder \cite{os} led to the establishment viewpoint that end state of a star, with remnant mass more than a few solar mass, would lead to formation of black holes in general theory of relativity. This model still serves as a role model for understanding some of the key features associated with black holes, appearance of singularity and event and apparent horizon formation \cite{hawking73}. It is no surprise that the phenomenon of visible singularity too was discovered many years later in dust collapse, generalising from homogeneous to inhomogeneous density \cite{smarr}. Today we have a complete understanding of the role of initial data in final state of collapse \cite{jdprd94}. Choice of dust as matter model has been supported for collapse studies since in late stages of continued collapse though we expect pressure to attain a upper bound the density continues to grow indefinitely \cite{penrose}.  Thus in higher dimensional generalisation of Einstein's theory of general relativity to Gauss-Bonnet gravity first we consider gravitational collapse of a spherically symmetric dust cloud.

The choice of dust as matter models makes comoving coordinates a preferred choice for a coordinate system. The line element in these coordinates can be written as
 \begin{equation}\label{metric}
 ds^2 = -dt^2 + A(r,t)^2 dr^2 + R(r,t)^2 d\Omega^2_{(d-2)}.
 \end{equation}
Here $d$ (5 or 6) is spacetime dimension. The only non-zero components of energy momentum tensor are
\begin{equation}
T^{\mu}_{\nu} = -\rho \delta^{\mu}_{t} \delta^{t}_{\nu}.
\end{equation}

\subsubsection{Collapse in five dimension}
The vanishing of $T^r_t$ component fixes
\begin{equation}
A(r,t)^2 = \frac{R_{,r}^2}{(1+f(r))}
\end{equation}
where, in analogy with four dimensional case, $f(r)$ can be identified as the total energy function. Considering special case of marginally bound dust, where dust particles start at rest from infinity, we can set $f(r)=0$. From (\ref{equation}) we write
\begin{eqnarray}
T^0_0 & = & -12\frac{{\dot R}^3 {\dot R}'}{R^3 R'} ,\nonumber \\
T^1_1 & = & -12\frac{{\dot R}^2 {\ddot R}}{R^3} ,\nonumber \\
T^{\theta}_{\theta} & = & -4\frac{{\dot R}^2 {\ddot R}'}{R^2 R'} - 8\frac{{\dot R} {\ddot R} {\dot R}'}{R^2 R'},\nonumber
\end{eqnarray}
and due to symmetry other angular components are equal to $T^{\theta}_{\theta}$. The vanishing of radial pressure gives two family of solutions, namely, static which correspond to ${\dot R} = 0$, and solutions with constant velocity
\begin{equation}
{\ddot R} = 0.
\end{equation}
Therefore,  area coordinate can be integrated as
\begin{equation}\label{solR}
R = r + b(r) t ,
\end{equation}
where $b(r)$ is an integration parameter. The other integration constant is fixed using $t=0$ as the scaling surface where $R=r$. It is straightforward to check that for this solution (\ref{solR}) other components of pressure vanish identically and we do not have any additional constraints.

The remaining freedom $b(r)$ can be interpreted as follows. Since we are considering marginally bound models where every collapsing shell is at rest at infinity. Thus $b(r)$ is identically zero in this case. Thus $R = r$, there is no collapse and we have a static solution. This is expected since gravity is kinematic in five dimension for Gauss-Bonnet gravity analogous to $3$-dimensional Einstein gravity.

\subsubsection{Collapse in six dimension}
As in five dimension, vanishing of $T^r_t$ component fixes
\begin{equation}\label{sol6}
A(r,t)^2 = \frac{R_{,r}^2}{(1+f(r))}.
\end{equation}
Again, considering special case of marginally bound dust we can set $f(r)=0$. From (\ref{equation}) we again have
\begin{eqnarray}\label{dens}
T^0_0 & = & -12\frac{{\dot R}^3}{{R^4} R'}[{\dot R}R' + 4 R {\dot R}'] , \\
T^1_1 & = & -12\frac{{\dot R}^2}{R^4}\left[{\dot R}^2 + 4R {\ddot R} \right],\\
T^2_2 & =& -12[R({\dot R}^2 {\ddot R})' + {\dot R}^2 ({\dot R} R')^{.}] \label{press}
\end{eqnarray}
The vanishing of radial pressure gives two family of solutions, namely, static which correspond to ${\dot R} = 0$, and solutions corresponding to
\begin{equation}
{\ddot R} = -\frac{{\dot R}^2}{4 R} .
\end{equation}
The first integration of equation above yields energy equation
\begin{equation}\label{eom}
{\dot R}^2 = \frac{2 m(r)}{R^{1/2}}.
\end{equation}
In the usual analogy with four dimensional case the left hand side acts as kinetic energy and right hand side plays the role of potential energy of the system. This allows us to define integration constant $m(r)$ as mass function. Note that unlike four dimensional case where potential energy goes as $1/R$ here it is $1/\sqrt{R}$. This equation can be integrated as
\begin{equation}\label{thesol}
R^{5/4} = r^{5/4} - \frac{5}{4} \sqrt{2 m(r)} t ,
\end{equation}
where we have used scaling freedom $(R(0,r) = r)$ to fix the integration constant. Again, the solution is strikingly similar to the four dimensional GR which can be obtained by replacing $5/4$ by $3/2$ everywhere. We could generalise it to any $2N+2$ dimension to write
\[
R^{(1+1/2N)} = r^{1+1/2N} -\frac{2N+1}{2N}\sqrt{2 m(r)} t.
\]
where, for we have GR, Gauss-Bonnet and cubic Lovelock respectively for $N = 1, 2, 3$ and so on.

From Eqs. (\ref{dens}) and (\ref{eom}) the expression for energy density is
\begin{equation}\label{dens2}
\rho(t,r) = 48 \frac{(m^2)'}{R^4 R'}.
\end{equation}
We have both shell crossing ($R'=0$) and shell focusing ($R=0$) singularities. It is interesting to draw a comparison with four dimensional case in Einstein gravity where we have derivative of the mass function in the numerator and not of its square. This arises because mass is a geometrical quantity and we have quadratic terms in curvature in the left hand side whereas right hand side is usual energy density. Generalising the argument for $2N+2$ dimensional marginally bound collapse, we write
\[
\rho(t,r) \propto \frac{(m^N)'}{R^{2N}R'}.
\]

The singularity curve in this case is given by
\begin{equation}\label{singu}
t_s(r) = \frac{4}{5}\frac{r^{5/4}}{\sqrt{2m(r)}}.
\end{equation}
It is instructive to calculate the time of formation of singularity at the centre.  The functional dependence of mass function near the centre can be calculated using Eq. (\ref{dens2}) as $m(r) \sim \rho_0 r^{5/2}/\sqrt{3} + \cdots $ where we have used scaling on the initial surface and $\rho_0$ is the initial central density of collapsing cloud. Thus time of formation of singularity at the centre is
\begin{equation}
t_s(0) \sim \frac{1}{\rho_0^{1/4}}.
\end{equation}
In the corresponding four dimensional Einstein gravity the time of singularity formation is inversely proportional to the square root of central density. Thus collapse slows down in Gauss-Bonnet theory. Moreover, in the naive dimensional extension of GR to six dimension the time of formation of central singularity continues to depend on central density as the inverse  square root. Extending to $2N+2$ dimensional case, we can write
\begin{equation}\label{gen-sing}
t_s(r) = \frac{2N}{2N + 1}\frac{r^{(2N +1)/2N}}{\sqrt{2m(r)}}
\end{equation}
and the central singularity in terms of central density as
\begin{equation}
t_s(0) \sim \frac{1}{\rho_{0}^{1/2N}}.
\end{equation}

To further our analogy of six dimensional Gauss-Bonnet gravity with four dimensional Einstein theory we consider now the causal structure of singularity. The  four dimensional inhomogeneous dust collapse model gave us the first generic violation of cosmic censorship conjecture \cite{jj}. In what follows we consider a simplified version of singular geodesic analysis by Barve et al. \cite{Barve}. To show if singularity is visible, at least locally, we consider out-going radial null geodesics in the six dimensional line element (\ref{metric}) with solution (\ref{sol6}) and $f(r)=0$ :
\begin{equation}
\frac{dt}{dr} = R' \;.
\end{equation}
If we assume existence of such geodesics with their past end-points at the central singularity $t_s(0)$, we can consider the following approximate form for the geodesics near the centre
\begin{equation}
t = t_s(0) + X r^{\delta}.
\end{equation}
Here both $\delta$ and $X$ are positive for geodesics to exist in spacetime. In case when leading exponent causing inhomogeneity is equal to $\delta$ the constant $X$ has a upper bound for singularity to be visible.

Consider an initial  density profile of the form
\begin{equation}\label{in-den}
\rho(0,r) = \rho_0 + \rho_n r^n
\end{equation}
to leading order near singularity. From Eq. (\ref{dens2}) we get the following form for the "mass function"  near the centre
\begin{equation}
4m(r)^2 = F_0 r^5 +F_n ^{n+5} \;.
\end{equation}
Now, using Eq. (\ref{thesol}), we can derive an expression for $R'$ near the centre, which we need for analysing radial null geodesics,
\begin{equation}
R' = \frac{\left[1-\frac{5}{4} {\cal F} t -n\frac{F_n}{4 F_0^{3/4}}r^n t  \right] }
{\left[1-\frac{5}{4} {\cal F} t\right]^{1/5}} ,
\end{equation}
where ${\cal F}=F_0^{1/4}(1+{F_n}r^n/{F_0}) $. Along the assumed singular geodesic we have $t = t_s(0) + X r^{\alpha}$, substituting $t$ and equating with $R'=dt/dr=X \alpha r^{\alpha  -1}$ we get the desired roots equation as
\begin{equation}\label{rootseqn}
X \alpha r^{\alpha  -1} = \frac{\left[1-\frac{5}{4} {\cal F} \tau_1 - n\frac{F_n}{4 F_0^{3/4}}r^n \tau_2  \right]}{
\left[1-\frac{5}{4} {\cal F} \tau_1\right]^{1/5}} ,
\end{equation}
where we  have defined $\tau_1=t_s(0)+Xr^{\delta}$ and $\tau_2=t_s(0)+Xr^{\alpha}$.
If this equation admits a solution with desired values of parameters we have at least one radial null geodesic terminating at singularity. As mentioned earlier for geodesics to lie in spacetime we should
have $\alpha \geq n$.  This in-equality can be easily understood since in case $\alpha < n$ the slope of outgoing radial null curves is more than that of singularity and thus are not part of spacetime.  In the first case $\alpha > n$ we have in the leading order
\begin{equation}\label{root1}
X \alpha r^{\alpha  -1} = \left(1+\frac{4}{5}n\right) \left(-\frac{F_n}{4F_0}\right)^{4/5} r^{4n/5}.
\end{equation}
Thus we have a self-consistent solution with $\alpha = 1 + 4n/5$ and $X = (-F_n/(4F_0))^{4/5}$.

Since we are considering the case $\alpha > n$, $n$ can take values 1 to 4. Again, this should be seen in comparison with what is observed in collapse of marginally bound dust in four dimensional Einstein gravity. When first non-zero term in the expansion of density around centre is either $\rho_1 <0$ or $\rho_2 <0$ singularity is always visible and black hole sets only for $n=3$ in four dimensional Einstein gravity. However in Gauss-Bonnet case this window enlarges it is only at $n=5$ that black hole sets in and it is naked singularity for all $\rho_1$ to $\rho_4$.

In case $\alpha=n$, the equation for the radial null curves can be written as
\begin{equation}\label{critroot}
nX r^{n-1} = \frac{\left[\left(1+\frac{4}{5}n \right)\left(-\frac{Fn}{4F_0}\right)-\frac{X}{t_s(0)}  \right]}{\left[\left(-\frac{F_n}{4F_0}\right)-\frac{X}{t_s(0)}\right]^{1/5}} r^{4n/5} .
\end{equation}
Thus $n=5$ and existence or otherwise of a positive definite root depends on the following sixth order polynomial equation,
\begin{eqnarray}\label{hex}
1280 Y^6 - \left[256 b+1\right] Y^5 + 5 b Y^4 -10 b^2 Y^3 \nonumber \\+10 b^3 Y^2 - 5 b^4 Y + b^5=0 .
\end{eqnarray}
Here we have defined $Y = X/F_0$ and $b = (-F_5/F_0^{9/4})$. Quantity $F_0$ is related to central density and hence is positive
definite, and since star should have a decreasing density away from the centre $F_5$ is negative and thus $b>0$. The parameter $b$ is a measure of inhomogeneity for a given central density. The existence
of a positive root depends on Eq.  (\ref{hex}) which is a sixth order polynomial and does not allow determination of roots in general. However, using Descartes' rule of signs this equation can admit six, four, two or zero (even) positive and no negative roots. Since we
have reduced it to a one parameter equation a simple numerical evaluation shows that for $0 < b < 0.000054$  this equation allows double roots. There is an additional restriction on $X$ by the slope of the singularity curve since we want geodesic lie in the spacetime. For $n>5$ we have $\alpha >n $ and we have only blackhole formation.

The generalisation to any dimension can be done as follows. If we consider a general density profile (\ref{in-den}), where $n$ characterises order of inhomogeneity, we can find an outgoing singular geodesic of the form $t = t_s(0) + X r^{\alpha}$ with constraint
$\alpha > n$ and $\alpha = 1 + 2N n/(2N+1)$ and the tangent $X\propto
(-F_n/F_0)^{2N/(2N+1)}$. Thus, as we move from GR to Lovelock more and more of initial data space, specified in terms of initial density, leads to formation of visible singularities. The threshold of transition is given by $n = 2N + 1$;i.e. $n=3, 5, ...$ for $N=1, 2, ...$ for corresponding to Einstein, Gauss-Bonnet and so on. The point to be noted is that there is a unique relation between inhomogeneity parameter $n$ and Lovelock order $N$ for the even dimension $d=2N+2$s.

The critical branch solution ($n=3$ in general relativity and $n=5$ in Gauss-Bonnet theory) serves as another distinguishing feature between naive extensions of GR to higher dimension and the Lovelock collapse. In 4d GR the well known transition between naked singularity and black  holes occurs at $n=3$ (for marginally bound case), where depending on other free parameters we can have either naked singularity or a black  hole. For $n<3$ singularity is always visible and for $n>3$ it is always covered. In five dimensional GR naked singularity window shrinks and it is naked only for $n=1$. For $n=2$ we have phase transition and for larger $n$ it is always a black hole. In six dimensional case this picture changes completely. For $n=1$ it continues to be visible whereas for any larger $n$ it is always a black hole. Therefore there is no more any transition branch, and the same is true for all higher dimensions than $6$. In contrast for Lovelock there would always occur transition phase for $n=2N+1$ in $2N+2$ dimension as demonstrated in particular for $6$-dimensional Gauss-Bonnet collapse. This is because gravitational (thereby collapse) dynamics is similar in all even $2N+2$ dimensions like the $4$-dimensional GR. Of course the window of visibility of singularity widens because Lovelock gravity becomes weaker with $N$ as potential goes as $1/r^{1/N}$.

The Lovelock curvature polynomial in six dimensions is given by,
\begin{equation}
R^{(2)}_{a b c d} = F^{(2)}_{a b c d} - \frac{1}{40} F^{(2)}(g_{ac} g_{bd} - g_{ad} g_{bc}),
\end{equation}
and the corresponding scalar invariant $K^{(2)} = R^{(2)}_{a b c d}  R^{(2) a b c d}$ reads
\begin{equation}
K^{(2)}_{(6)}  = 135\frac{m^4}{R^{10}}  - 108 \frac{m^3 m'}{R^9 R'}
+144 \left(\frac{m m'}{R^4 R'}\right)^2.
\end{equation}
It is instructive to compare it with the corresponding six dimensional GR case. The acceleration equation corresponding to vanishing of radial pressure is given by
\begin{equation}
{\ddot R} = -\frac{3 {\dot R}^2}{2R}
\end{equation}
and the Kretschmann scalar $K^{(1)}=R_{a b c d}  R^{a b c d}$ is
\[
K^{(1)}_{(6)} = 240\frac{m^2}{R^{10}}  - 96 \frac{m m'}{R^9 R'}
+13 \left(\frac{m'}{R^4 R'}\right)^2 .
\]
The Kretschmann scalar in either case has the similar fall off behavior. This can be understood as follows. For $d=2N+2$, Einstein potential goes as $1/R^{2N-1}$ and curvature as $1/R^{2N+1}$ and hence $K^{(1)}$ will go as $1/R^{2(2N+1)}$ while for Lovelock we have respectively $1/R^{1/N}$, $1/R^{(2N+1)/N}$ and $(1/R^{(2N+1)/N})^{2N} = 1/R^{2(2N+1)}$. The Kretschmann scalar has therefore the same degree of divergence for both Einstein and Lovelock gravity. This shows that strength of singularity remains the same and it does not distinguish between Einstein and pure Lovelock gravity for the given even dimension. However for Lovelock gravity there is no collapse in any odd dimension.

\subsection{Null dust collapse}
The gravitational collapse of spherical matter in the form of
radiation (null dust) described by the Vaidya metric is
well studied \cite{psj}.   In GR,   it turns out that as dimension increases, the
window for naked singularity shrinks. That is, gravity seems
to get strengthened with an increase in dimensions of space \cite{gd}.
In this context,
one question that could naturally arise is, what happens in
gravitational collapse of null dust in the pure Lovelock gravity?

Inclusion of Vaidya null dust in said contest is quite straightforward by writing the corresponding vacuum solution in Eddington advanced time and then making mass function of it. It would then describe a null radiation zone with radially flowing null rays.
Let us begin with the general $d$-dimensional spherically
symmetric spacetime, in advanced Eddington time coordinate $v$,
described by the metric \cite{pd,ghd}:

\begin{equation}
ds^2 =  f(v,r)  dv^2\; + 2 dv dr  + r^2 d {\Omega}_{d-2}^2
\label{eq:me}
\end{equation}
where $0 \leq r \leq \infty$ is the proper radial coordinate,
$-\infty \leq v \leq \infty$ is an advanced time coordinate, and
\begin{eqnarray}
d \Omega_{d-2}^2 &=& d \theta^2_{1} + \sin^2({\theta}_1) d
\theta^2_{2} + \sin^2({\theta}_1) sin^2({\theta}_2)d \theta^2_{3}
\nonumber \\ & & + \ldots + \left[\left( \prod_{j=1}^{d-3}
\sin^2({\theta}_j) \right) d \theta^2_{d-2} \right].
\end{eqnarray}
It proves useful to introduce a local mass function $m(v,r)$ defined by
$f(v,r)=1 - 2m/r$ \cite{bi}. Here $m(v,r)$ is an arbitrary function
of advanced time $v$ and radial coordinate $r$. When $m = M(v)$, it
is the Vaidya solution in higher dimension \cite{pd,gd}. The usual Vaidya solution
follows for  $d=4$ and $m = M(v)$.

For obtaining analogue of Vaidya solution \cite{pc} in Lovelock gravity, we note that $T_{\mu\nu} = \sigma k_\mu
k_\nu, ~k_\mu k^\mu = 0$ where $\sigma$ is density of null
fluid. The equation we need to solve is
\begin{equation}
G^{(N)}_{ab}=\Lambda g_{ab} + \sigma k_\mu k_\nu  \label{eq:gmn}
\end{equation}
which is Lovelock gravitational equation. The standard way to do it is to transform Lovelock analogue of Schwarzschild solution to Eddington advanced time and write $m=M(v)$. This would satisfy the above equation and the metric would read as follows:
\begin{eqnarray}
ds^2 = - \left[1 - \left(\Lambda r^{2N} +
\frac{M(v)}{r^{d-2N-1}}\right)^{1/N}\right] dv^2 \nonumber \\+2 dv dr + r^2
d\Omega_{d-2}^2. \label{solA}
\end{eqnarray}
This is Lovelock analogue of Vaidya solution for $d=2N+2$. It would automatically match with the static version at the null boundary. In odd $d=2N+1$, we have Lovelock analogue BTZ black hole which would exist in all odd dimensions.

We now turn to collapse of null dust to see under what conditions black hole or  naked singularity result? The physical situation here is that of a radial influx of null fluid in an initially empty region of Lovelock de Sitter spacetime. The first shell arrives at $r=0$ at time $v=0$ and the final at $v=T$. A central singularity of growing mass is developed at $r=0$. For $ v < 0$ we have $m(v)\;=\;0$,i.e.  Lovelock de Sitter spacetime, and for $ v > T$, $\dot{M}(v)\;=\;0$, $M(v)\;$ is positive definite. The metric
for $v=0$ to $v=T$ is Lovelock Vaidya, and for $v>T$ we have Lovelock Schwarzschild solution. In order to get an analytical solution
for our case, we choose,
\begin{equation}
M(v) = \left\{ \begin{array}{ll}
    0,          &   \mbox{$ v < 0$}, \\
    \lambda v^{(d-2N-1)} \; (\lambda>0)
                                &   \mbox{$0 \leq v \leq T$}, \\
    m_{0}(>0)       &   \mbox{$v >  T$}.
        \end{array}
    \right.             \label{eq:mv}
\end{equation}

For radial null geodesics, we readily write
\begin{equation}
\frac{dr}{dv} = \frac{1}{2} \left[1 - \left(\Lambda r^{2n} +
\frac{M(v)}{r^{d-2N-1}}\right)^{1/N}\right].
\label{eq:de1}
\end{equation}
Clearly, this has singularity at
$r=0$, $v=0$. The nature (naked singularity or black hole) of
the collapsing solutions can be characterized by the existence of
radial null geodesics coming out of
singularity. The motion near singularity is characterized by roots
of an algebraic equation which we derive next.
Eq. (\ref{eq:de1}), upon using  eq. (\ref{eq:mv}), turns out to be
\begin{equation}
\frac{dr}{dv} = \frac{1}{2} \left[1 - \left(\Lambda r^{2N} +
{X^{d-2N-1}}\right)^{1/N}\right]
\label{eq:de2}
\end{equation}
where $X \equiv v/r$ is the tangent to a possible outgoing geodesic.
The central shell focusing
singularity is at least locally naked
(for brevity we have addressed it as
naked throughout this paper), iff there exists $ X_0 \in (0, \infty )$
which satisfies
\begin{eqnarray}
 X_{0} &=&  \lim_{r \rightarrow 0 \; v\rightarrow 0} X =
\lim_{r\rightarrow 0 \; v\rightarrow 0} \frac{v}{r}= \nonumber \\&&
\lim_{r\rightarrow 0 \; v\rightarrow 0} \frac{dv}{dr} =
\frac{2}{1 - \left(
{X^{d-2N-1}}\right)^{1/N}}   \label{eq:lm1}
\end{eqnarray}
or,
\begin{equation}
\lambda X_{0}^{m} - X_{0} + 2 = 0,      \label{eq:ae}
\end{equation}
with $m=(d-N-1)/N$. Thus any solution $X= X_0 > 0$ of the eq. (\ref{eq:ae}) would
correspond to
naked singularity of spacetime;i.e. to future directed
null geodesics emanating from singularity  $(v=0, r=0)$.
The smallest such $X_0$ corresponds to the earliest ray
emanating from singularity and is called Cauchy horizon of spacetime. If $X_0$ is the smallest positive root of  (\ref{eq:ae}), then there are
no naked singularities in the region $X < X_0$.
Hence in the absence of positive real roots, the central
singularity is not naked (censored) because in that case there are
no outgoing future directed null geodesics emanating from singularity.
Thus,occurrence of positive real roots implies that the strong CCC is violated, though not necessarily the weak CCC.

Obviously for the critical dimension $d=2N+1$, we get $m=1$ and Eq.~(\ref{eq:ae}) has a trivial solution $X_0=2/(1-\lambda)$. This is as expected as gravity is kinematic here.

We now examine the condition for occurrence of naked singularity for $d=2N+2$.
With a straightforward calculation it can be shown that eq.
(\ref{eq:ae}) always admits two real positive roots $X_1$ and $X_2$ for $\lambda
\leq \lambda_c$,  where $\lambda_c$ is the critical value of the
parameter $\lambda$ discriminating between naked singularity and
black hole. The larger of the two roots, say $X_2$ corresponds to Cauchy horizon. Thus it follows that singularity will be naked if $\lambda \leq \lambda_c$. On the other hand if, the inequality is reversed, $\lambda > \lambda_c$ no naked singularity would form and
gravitational collapse would result in black hole. When  $\lambda = \lambda_c$, the two roots coincide to $X_C$ (say), and then from
Eq.~(\ref{eq:ae}), we have
\begin{equation}
\lambda\; m\; X_C^{m-1} -1 =0,\; \mbox{i.e.,}\; X_C = \left(\frac{1}{\lambda m} \right)^{\frac{1}{m-1}}
\end{equation}
and which on substituting into (\ref{eq:ae}), results to
\begin{equation}
\lambda_C=\frac{1}{m} \left( \frac{m-1}{2m}\right)^{(m-1)} = \frac{N}{N+1}(2(N+2))^{-1/N}
\end{equation}
and the value of the critical root at $\lambda_C$
\begin{equation}
X_C=\frac{2m}{m-1} = 2(N+1).
\end{equation}
For the familiar Einstein case $N=1$, we have $\lambda_C = 1/8$ and $X_C = 4$ while they are $\lambda_C=1/3\sqrt{2}$ and $X_C = 6$ for $N=2$ Gauss-Bonnet gravity.

\subsubsection {Kretschmann Scalar}

For gauging strength of singularity we compute Kretschmann scalar which is square of corresponding curvature and it shows how curvature diverges. Let us set $\Lambda=0$ as it represents constant curvature. Lovelock gravity is always kinematic in odd dimension which will give null $K^{(2)}$ while it will not be Riemann flat and $K^{(1)}$ will be non-zero falling off as $1/r^4$.

For the dynamic $6$-dimensional case, we have
\begin{equation}
K^{(2)}_{(6)} = 135\frac{M(v)^2}{r^{10}}
\end{equation}
while its corresponding Riemann square goes:
\begin{equation}
R_{a b c d}  R^{a b c d} = \frac{425 M(v)}{16 r^5}.
\end{equation}
However the corresponding scalar for $6$ dimensional Einstein collapse is given by
\begin{equation}
K^{(1)}_{(6)} = \frac{240 M(v)^2}{ r^{10}}.
\end{equation}
which has the same fall off as Gauss-Bonnet collapse. This is rather remarkable that the strength of the singularity does not discriminate between Einstein and Lovelock gravity.

\section{Discussion}

In Einstein gravity for dust (null as well as non-null) collapse, it turns out that as dimension of spacetime is increased the threshold inhomogeneity separating black hole and naked singularity also increases. That is compared to $4$ dimension, greater degree of inhomogeneity is required for formation of naked singularity. Higher dimension therefore favors black hole. In contrast, for pure Lovelock gravity, it is the other way round. As $d=2N+2$ increases threshold decreases thereby implying even a lesser degree of inhomogeneity could lead to naked singularity formation. In contrast to Einstein, Lovelock gravity hence favors naked singularity as even dimension increases. This could be understood as follows. In Einstein gravity, gravitational force goes as $1/r^{d-2}$ which means it becomes stronger with increase in dimension. Stronger gravitational pull  should naturally favor black hole. On the other hand for Lovelock, force goes as $1/r^{(N+1)/N} = 1/r^{d/(d-2)}$ and that clearly weakens with $N$ or $d$. That is why it favors naked singularity.

Despite this contrasting behavior, the Kretshmann scalars for both Einstein and Lovelock however have the same order of divergence indicating the same singularity strength. This happens because for Lovelock potential goes as $1/r^{1/N}$, Riemann will go as $1/r^{2+1/N}$ and the corresponding Kretshmann scalar as
$(1/r^{(2N+1)/N})^{2N} = 1/r^{2(2N+1)}$ while for Einstein they go respectively as $1/r^{d-3}$, $1/r^{d-1}$ and $1/r^{2(d-1)}$ (for $d=2N+2$, $d-1=2N+1$). Thus even though for a given degree of inhomogeneity, collapse may result in a black hole for Einstein and naked singularity for Lovelock, yet the Kretshmann scalar for both would have the same order of divergence. This shows that rate of collapse which determines black hole or naked singularity depends upon Einstein or Lovelock but the ultimate order of divergence of singularity is neutral.

\section{Acknowledgments}   SGG would like to thank
University Grant Commission (UGC) for major research project grant NO. F-39-459/2010(SR) and to IUCAA, Pune for kind hospitality while part of this work was being done.
SJ acknowledges support under UGC minor research project (42-1068/2013(SR).

\bibliographystyle{elsarticle-num}

\end{document}